\documentclass{iopconfser}
\usepackage{amsmath}
\usepackage{graphicx}
\begin{document}

\title{Effects of Symmetron on growth and RSD multipoles}

\author{Gerardo Morales-Navarrete$^{1,2}$, Jorge L.~Cervantes-Cota$^{1}$ }

\affil{$^1$Departamento de F\'isica, Instituto Nacional de Investigaciones Nucleares, Apartado Postal 18-1027, Col. Escand\'on, Ciudad de M\'exico,11801, M\'exico. \\
$^2$Instituto de F\'isica, Universidad Nacional Aut\'onoma de M\'exico, Apartado Postal 20-364, Ciudad de M\'exico, M\'exico.}

\email{moralesnavarretegerardo@gmail.com, jorge.cervantes@inin.gob.mx}

\begin{abstract}
In this work, we investigate the effects of the growth rate scale dependence in the Symmetron modified gravity (MG) model on cosmic structure formation and we analyze the redshift-space distortion (RSD) multipoles, comparing with the Hu-Sawicki $f(R)$ model (specifically the F6) and the standard $\Lambda$CDM model. The analysis employs a scale-dependent growth equation and utilizes the fk-PT perturbation theory approach, implemented in the FOLPS-nu code, to compute the full 1-loop power spectrum multipoles, in particular, the monopole and quadrupole ($\ell=0,2$, respectively). The results show that at redshift $z=0$, the monopole of both MG models is suppressed compared to $\Lambda$CDM, with the Symmetron being closer to the standard model, while the quadrupole presents the opposite behavior. To validate the pipeline, we use  General Relativity (GR) mock catalogs (EZMocks), since suitable Symmetron simulations are not available. The main result is that the Markov Chain Monte Carlo (MCMC) analysis successfully recovers the expected GR limit (i.e., $\beta_0 \approx 0$) from the Symmetron model when applied to this mock data, confirming the viability of our methodology for cosmological inference. Then, we conclude that the pipeline is prepared to test MG models against current and near-future galaxy surveys.
\end{abstract}

\section{Introduction}
The standard model of cosmology, $\Lambda$CDM model, is the more successful model that explains observations such as the cosmic microwave background and baryon acoustic oscillations. However, it relies on the cosmological constant as a source for the expansion of the Universe, a hypothesis that faces theoretical challenges. Furthermore, recent data from surveys like DESI and DES now suggest that models concerning dynamical dark energy (such as wCDM or $\text{w}_0 \text{w}_a$CDM) provide a better explanation for the observations, showing a tension with the $\Lambda$CDM model at a significance level of up to $4.2\sigma$.

Regarding the above, there is a new interest in exploring alternative theories, specifically Modified Gravity (MG) models, which can explain dark energy in a physical form rather than a simple parametrization. Also, with the new data from current and future galaxy surveys, it is possible to test these models by analyzing cosmological observables to understand the large-scale structure, in particular through the power spectrum. In this work, we analyze two MG models, the Symmetron and Hu-Sawicki $f(R)$ models, that incorporate screening mechanisms to remain consistent with solar system tests while predicting distinct clustering properties that can be tested with modern and efficient tools and upcoming data. To see details, read the work \cite{moralesnavarrete2025fullshapepowerspectrumsymmetron}.
\section{Growth rate scale dependency}
To analyze the density contrast growing mode ($D_+$), we use the growth rate ($f \equiv d \text{ln} D_+/d \text{ln} a$) in terms of the e-folds as the time variable, which is convenient for the implementation of the code FOLPS-nu \cite{Noriega:2022nhf}. It turns out that the expression for general cosmologies, beyond $\Lambda$CDM, is \cite{Linder_2005}:
\begin{eqnarray} \label{f_MG}
    f^{'} + f^{2} + \left[ 2 - \frac{3}{2}  \left ( 1 + w_\phi (1- \Omega_m - \Omega_\Lambda) + w_{\Lambda} \Omega_{\Lambda} \right)  \right] f - \frac{3}{2} \Omega_m \mu(k,a)  = 0,
\end{eqnarray}
where $\frac{d}{dN} \equiv (~)'$, $N= \ln(a)$, $\omega_\phi$ is the parameter of the equation of state associated with the scalar field, $\Omega_m = \rho_m(t) / \rho_c (t)$ with $\rho_c (t) = 3H^2(t) /8\pi G$,  and $w_{\Lambda}$ is the dark energy equation of state that we take as $\Lambda$CDM, $w_{\Lambda}=-1$, which acts as a friction term for the growth function equation, cf. \cite{moralesnavarrete2025fullshapepowerspectrumsymmetron}. Eq. (\ref{f_MG}) is valid for MG that is characterized by the $\mu(k,a)$ function, which leads to an enhancement of the effective matter density's contrast and its growth rate. These terms imply a change in the structures' collapse in comparison to $\Lambda$CDM. For MG, the source of the Poisson equation depends on scale through the function $\mu(k,a)$, the growth functions and growth rate inherit this property too. Nevertheless, there exists a complication due to the scale dependence that slows the computations and  makes the comparison with the measured growth rate inefficient. 

Here, we use the fk-PT approach, which consists of taking the $\Lambda$CDM kernels, while keeping the scale dependence in the linear growth rate. As shown in \cite{Noriega:2022nhf} this approximation is useful for the $\Lambda$CDM plus neutrinos, and it also works for HS-$f(R)$ as is shown in \cite{Rodriguez_Meza_2024}. To see some $f(k,z)$ and other plots, we refer the reader to the work \cite{moralesnavarrete2025fullshapepowerspectrumsymmetron}.
\section{1-loop RSD MG multipoles}
We compute the RSD multipoles for the Symmetron model, comparing against $n=1$ HS-$f(R)$ and taking $\Lambda$CDM as a reference model.
To do these calculations, we use Eqs. (4.44) and (4.48) from Ref. \cite{moralesnavarrete2025fullshapepowerspectrumsymmetron}, which take into account the full 1-loop approximation that is implemented in the code FOLPS-nu, which is normally valid for $\Lambda$CDM plus neutrinos. Accordingly, we implement the two MG models studied in \cite{moralesnavarrete2025fullshapepowerspectrumsymmetron}: A corrected version for the HS-$f(R)$ and Symmetron.
\begin{figure}[ht]
 	\begin{center}
  \includegraphics[width=3.0 in]{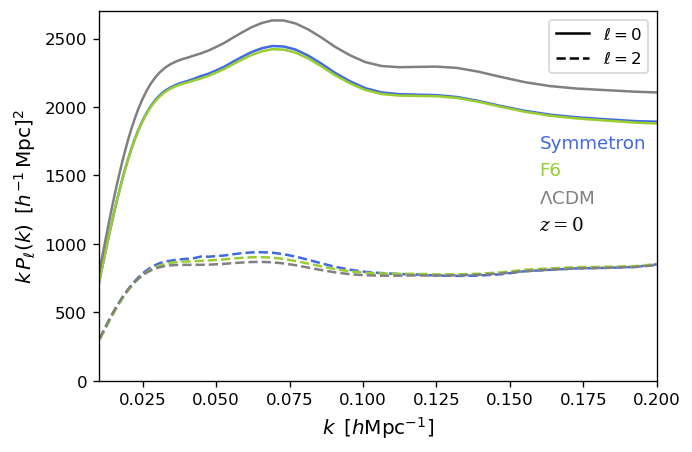}
  \includegraphics[width=3.0 in]{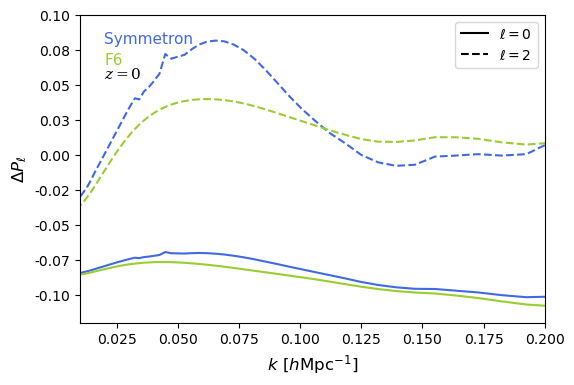}
 	\caption{Monopole ($\ell=0$) and quadrupole ($\ell=2$) at $z=0$ for the Symmetron, Hu-Sawicki F6, and $\Lambda$CDM models.}
    \label{fig:Pl02h12z05}
 	\end{center}
\end{figure}
The cosmological parameters used were $\{ \Omega_m, \Omega_b, h, n_s \}=$ $\{ 0.281, 0.046, 0.697, 0.971 \}$ and nuisance parameters: EFT counterterms $\{ \alpha_0, \alpha_2, \alpha_4, \tilde{c} \}$, bias $\{ b_1, b_2, b_{s^2}, b_{3nl} \}$, and shot noise terms. As is common, we use the co-evolution model \cite{Chan_2012,Baldauf_2012,Saito_2014} to reduce the independent parameters from 8 to 6. We show one of our results in Fig. \ref{fig:Pl02h12z05} ( at $z=0$), where a gray line corresponds to $\Lambda$CDM, a green line corresponds to F6 model, and the blue is for Symmetron. For this plot, we have set the three MG parameters of Symmetron, $\beta_0=1$, $m_0=1$, $a_{ssb}=0.5$, and we fix the $\Lambda$CDM background behavior for values $a \leq a_{ssb}$. On the right panel
\begin{eqnarray}
    \Delta P_\ell &=& (P_\ell^{MG} - P_{\ell}^{GR} )/P_{\ell}^{GR},
\end{eqnarray}
is the deviation on the MG multipole with respect to $\Lambda$CDM. We can see in Fig. \ref{fig:Pl02h12z05} that the monopole for F6 and Symmetron is  smaller than $\Lambda$CDM, with Symmetron being closer to $\Lambda$CDM at $z=0$. In \cite{moralesnavarrete2025fullshapepowerspectrumsymmetron} other plots are shown at $z=0.5$ and $z=1$ for $k<0.25 ~h~ \text{Mpc}^{-1}$. The quadrupole exhibits an opposite behavior, in all cases, F6 and Symmetron are larger than $\Lambda$CDM for these redshifts, as we can see here too.
\section{Model validation with GR mocks}
The implementation of MG models within the FOLPS-nu code utilizing the fk-PT approximation and the Infrared Resummation (IR-Resummation) requires validation against simulations. However, existing simulations for the Symmetron model are not suitable for this purpose, as they either focus on studying the linear matter power spectrum or only on code development rather than high-precision constraints, making them ineffective for testing modeling like 1-loop redshift space distortions or the IR-resummation scheme. 
\begin{figure}[ht]
 	\begin{center}
 	\includegraphics[width=3.8 in]{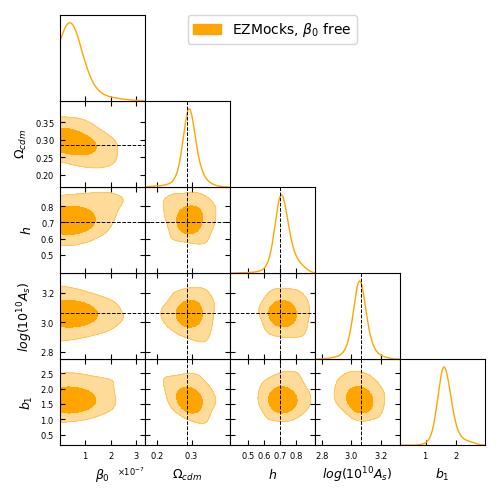}
 	\caption{Confidence level contours for Symmetron model using EZMocks at $z=0$, using the monopole and quadrupole. The $\beta_0$ parameter was considered free with a flat prior $\mathcal{U}(-0.01,0.01)$ as similar as $f_{R0}$ for $f(R)$ in Appendix B of Ref. \cite{moralesnavarrete2025fullshapepowerspectrumsymmetron}. Dashed lines mark EZMocks parameter simulation values.}
 	\label{fig:contour_symm}
 	\end{center}
\end{figure}
Therefore, we use General Relativity mocks (EZMocks) to validate the MG pipeline and verify that it can correctly recover GR parameters in the MCMC analysis. The result of this validation is presented in Fig. \ref{fig:contour_symm} showing the confidence level of fitting the Symmetron model. The aim was to recover the correct GR behavior from the Symmetron, which is represented by a value of zero for the MG coupling parameter $\beta_0$, and where the estimated value is less than $10^{-7} $ at $65\%$ of confidence level. The analysis successfully achieved this, with the estimated parameters showing a good fit to the mock data and the contours centering on the expected $\Lambda$CDM values, as indicated by the dashed lines. This successful recovery of GR from the Symmetron framework, using the mean of seven mock realizations, confirms the robustness and reliability of the implemented pipeline for cosmological parameter inference with real survey's data.
\section*{Acknowledgments}

The authors thank the support by SECIHTI (CONAHCyT) project CBF2023-2024-589. Gerardo Morales-Navarrete thanks the SECIHTI (CONAHCyT) for the financial support provided through the research assistant grant (C-823089/2024), under the SNII program.
 \bibliography{growth_MG.bib}  
\end{document}